\documentclass{elsart}
\usepackage{natbib}
\usepackage{epsfig}
\usepackage{amssymb}

\begin{document}

\begin{frontmatter}

\title{Difference in nature of correlation between NASDAQ and BSE indices}

\author{P. Manimaran}
\address{School of Physics, University of Hyderabad,
 Hyderabad 500 046, India}
\author{Prasanta K. Panigrahi$^{1,2}$, and}
\ead{prasanta@prl.res.in}
\author{Jitendra C. Parikh$^2$}
\address{$^1$Indian Institute of Science Education and Research (Kolkata), Salt Lake,
Kolkata 700 106, India}
\address{$^2$Physical Research Laboratory, Navrangpura, Ahmedabad
380 009, India}

\begin{abstract}
We apply a recently developed wavelet based approach to characterize
the correlation and scaling properties of non-stationary financial
time series. This approach is local in nature and it makes use of
wavelets from the Daubechies family for detrending purpose. The
built-in variable windows in wavelet transform makes this procedure
well suited for the non-stationary data. We analyze daily price of
NASDAQ composite index for a period of 20 years, and BSE sensex
index, over a period of 15 years. It is found that the long-range
correlation, as well as fractal behavior for both the stock index
values differ from each other significantly. Strong non-statistical
long-range correlation is observed in BSE index, whose removal
revealed a Gaussian random noise character for the corresponding
fluctuation. The NASDAQ index, on the other hand, showed a
multifractal behavior with long-range statistical correlation.

\end{abstract}

\begin{keyword}
Time series \sep fluctuations \sep fractals \sep discrete wavelets
\sep Hurst exponent

\PACS 05.45.Tp \sep 89.65.Gh \sep 05.45.Df \sep 52.25Gj
\end{keyword}

\end{frontmatter}

\section{Introduction}

Non-stationary time series have been investigated through several
approaches. In particular, the characterization of fluctuations and
their scaling behavior have been the focus of many studies, since
they reveal the nature of the dynamics. In this context, financial
time series have attracted considerable attention \cite{Mante}. The
large length of the available data of various stock market indices
make them ideal candidates for analysis. Furthermore, the complex
dynamics of the variations in stock prices yield fluctuations which
can show correlations, as well as scaling behavior. The goal of the
present paper is to study the self-similar and correlation
properties of NASDAQ and BSE indices belonging to two different
economical environments. These stock indices belonging to developed
and developing countries may show characteristic differences,
possibly arising due to differences in their underlying dynamics. We
concentrate on the nature of correlations and fractal behavior for
which wavelet transform \cite{daub,mall} is used as a tool for
extracting the fluctuations at different scales.

A number of methods have been devised to find scaling behavior in
time series. The well-known structure function method \cite{ba} and
the recently developed wavelet transform modulus maxima (WTMM)
method \cite{arn1}, relying on continuous wavelet transforms, are
widely used for the analysis of stationary data. The fact that most
of the time series arising in real systems are non-stationary in
nature introduces complications in estimating the scaling behavior,
while using the above two approaches, which are global in nature.
Hence, in recent times, local approaches, like detrended fluctuation
analysis (DFA) and its generalization MF-DFA
\cite{gopi,muzy,pen,khu,jan} have been developed to handle
non-stationary data. In this case, one uses windows of various sizes
to separate fluctuations from the trend, which can also be shuffled
to remove any correlation in the data set. While isolating the
average or the trend of the data points in a given window, one takes
recourse to linear or quadratic fit in the DFA approach. We have
introduced a new method based on discrete wavelets
\cite{mani1,mani2,ran} to characterize the scaling behavior of
non-stationary time series. The present procedure is similar to
those in MF-DFA \cite{jan}, except that in order to detrend, we use
wavelets and MF-DFA uses local polynomial fits. Recently, this
method has been used by Brodu, to analyze in real time, fractal
behavior of dynamic time series \cite{nico}. The relative merits of
MF-DFA and a variety of other approaches to characterize
fluctuations have been carried out in Ref.\cite{jaro}. It is worth
emphasizing that fluctuation analysis and characterization have been
earlier attempted using Haar wavelets, in the context of bio-medical
applications, without the study of scaling behavior
\cite{pkp1,pkp2}.

Wavelets from Daubechies family are used for extracting trend from
the given data set. Fluctuations are captured by high-pass
coefficients and the trend captured by the low-pass coefficients of
wavelet transform. The discrete wavelet transform provide a handy
tool for isolating the trend in a non-stationary data set, because
of its built-in ability to analyze data in variable window sizes. In
this note, we analyze returns of stock index values through our new
wavelet based method. Multi-fractal properties are also investigated
using multi-fractal spectrum. We analyze daily price of NASDAQ
composite index for a period of 20 years, starting from 11-Oct-1984
to 24-Nov-2004, and BSE sensex index, over a period of 15 years,
starting from 2-Jan-1991 to 12-May-2005.

\section{Data analysis}
It is found that the nature of correlation is quite different
between these two financial time series and significant
non-statistical correlation exists in both of them. Removal of the
same reveals that BSE index is primarily mono-fractal with the
fluctuations showing a Gaussian random noise character. On the other
hand, the NASDAQ index shows a weak multifractal behavior with long
range statistical correlation.

\begin{figure}
\centering
\includegraphics[width=3in]{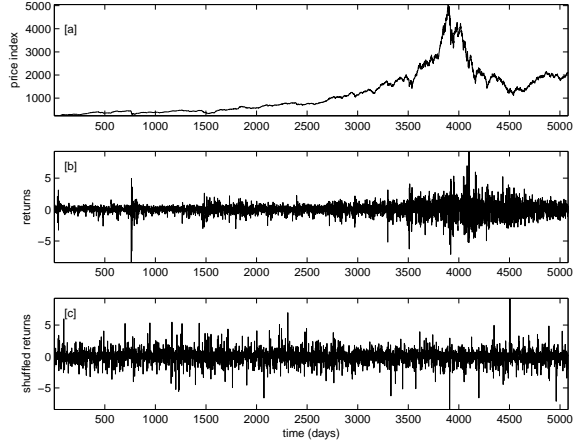}
\caption{[a] NASDAQ daily (close) composite index for a period of 20
years, starting from 11-Oct-1984 to 24-Nov-2004, [b] daily returns
show more clusters of small and large fluctuations and [c] the
returns after shuffling show disappearance of clustering behavior.}
\end{figure}

\begin{figure}
\centering
\includegraphics[width=3in]{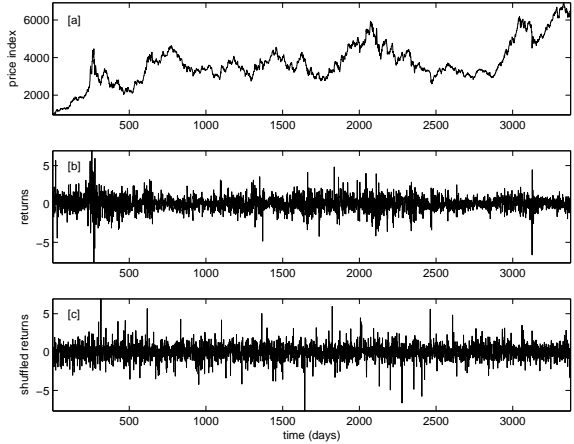}
\caption{[a] BSE sensex daily (close) index value for a period of 15
years, starting from 2-Jan-1991 to 12-May-2005, [b] daily returns
show much less (compared to Fig. 1 (b)) clustering of fluctuations
and [c] the returns after shuffling show no significant difference
in appearance relative to Fig. 2 (b).}
\end{figure}

From the financial (NASDAQ composite index and BSE sensex) time
series $x(t)$, we first compute the scaled returns defined as,

\begin{equation}
G(t)\equiv [ln(x(t+1))- ln(x(t))]/\sigma,~~~ t=1,2...(N-1);
\end{equation}

here $\sigma$ is the standard deviation of $x(t)$. From the returns,
the signal profile is estimated as the cumulative,
\begin{equation}
Y(i) = \sum_{t=1}^i [G(t)], ~~~ i=1,....,N-1.
\end{equation}

Next, we carry out wavelet transform on the profile $Y(i)$ to
separate the fluctuations from the trend by considering precise
values of window sizes $W$ corresponding to different levels of
wavelet decomposition. We obtain the trend by discarding the
high-pass coefficients and reconstructing the trend using inverse
wavelet transform.  The fluctuations are then extracted at each
level by subtracting the obtained time series from the original
data. Though the Daubechies wavelets extract the fluctuations
nicely, its asymmetric nature and wrap around problem affects the
precision of the values. This is corrected by applying wavelet
transform to the reverse profile, to extract a new set of
fluctuations. These fluctuations are then reversed and averaged over
the earlier obtained fluctuations. These are the fluctuations (at a
particular level), which we consider for analysis. In Figs. 1 and 2,
we give the time series for the two index data sets, and the
corresponding returns. We also show shuffled returns for the two
series to examine the correlation as well as "bursty" (clustering)
behaviour.

The extracted fluctuations are subdivided into non-overlapping
segments $M_s = int(N/s)$ where $s=2^{(L-1)}W$ is the wavelet window
size at a particular level $L$ for the chosen wavelet. Here $W$ is
the number of filter coefficients of the discrete wavelet transform
basis under consideration. For example, with Db-4 wavelets, $s=4$ at
level 1 and $s=8$ at level 2 and so on. It is obvious that some data
points would have to be discarded, in case $N/s$ is not an integer.
This causes statistical errors in calculating the local variance. In
such cases, we have to repeat the above procedure starting from the
end and going to the beginning to calculate the local variance. The
detrending and extracted fluctuations have been depicted in Figs 3
and 4.

\begin{figure}
\centering
\includegraphics[width=3in]{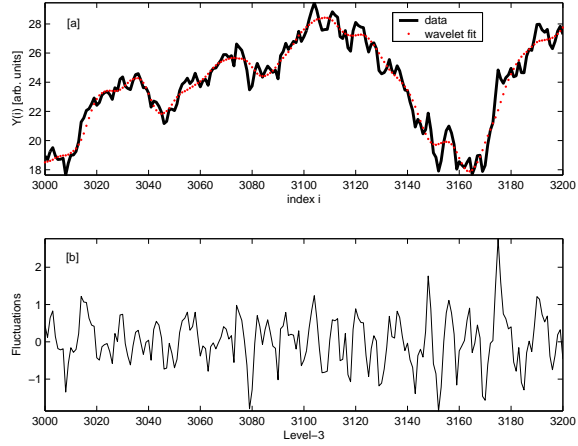}
\caption{[a] Detrending the integrated returns of NASDAQ composite
index at the scale level-3, through Db-8, wavelet, and [b] the
extracted fluctuations at the scale level-3 (window size 32).}
\end{figure}

\begin{figure}
\centering
\includegraphics[width=3in]{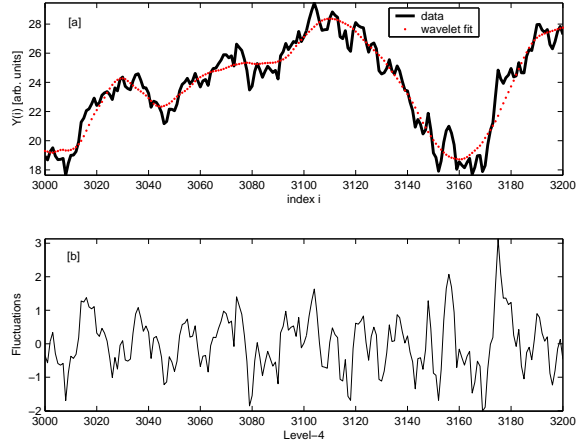}
\caption{[a] Detrending the integrated returns of NASDAQ composite
index at the scale level-4, through Db-8, wavelet, and [b] the
extracted fluctuations at the scale level-4 (window size 64).}
\end{figure}

The $q^{th}$ order fluctuation function $F_q(s)$ is obtained by
squaring and averaging fluctuations over all segments:
\begin{equation}
F_q(s) \equiv  \{ \frac {1}{2 M_s} \sum_{b=1}^{2 M_s} [
F^2(b,s)]^{q/2}\}^{1/q}.
\end{equation}

Here '$q$' is the order of moments that takes real values. The above
procedure is repeated for variable window sizes for different values
of $q$ (except $q=0$). The scaling behaviour is obtained by
analyzing the fluctuation function,

\begin{equation}
F_q(s) \sim s^{h(q)},
\end{equation}
in a logarithmic scale for each value of $q$. If the order $q = 0$,
direct evaluation through Eq. (3) leads to divergence of the scaling
exponent. In that case, logarithmic averaging has to be employed to
find the fluctuation function:

\begin{equation}
F_q(s) \equiv exp \{ \frac {1}{2 M_s} \sum_{b=1}^{2 M_s} ln [
F^2(b,s)]^{q/2} \}^{1/q}.
\end{equation}

\section{Results and Discussion}
As is well-known, if the time series is monofractal, the $h(q)$
values are independent of $q$. For multifractal time series the
$h(q)$ values depend on $q$. The correlation behaviour is
characterized from the Hurst exponent ($H=h(q=2)$), which varies
from $0 < H < 1$. For long range correlation, $H > 0.5$, $H=0.5$ for
uncorrelated and $H <0.5$ for long range anti-correlated time
series.

The scaling exponent is calculated for various values of $q$ for
both the stock index values. Figs. 5 and 6, show the way $h(q)$ and
$\tau(q)$ vary with $q$ for the returns and shuffled returns of the
two time series. The non-linear behaviour of $h(q)$ for different
$q$ values, is the measure of multifractality.

\begin{figure}
\centering
\includegraphics[width=3in]{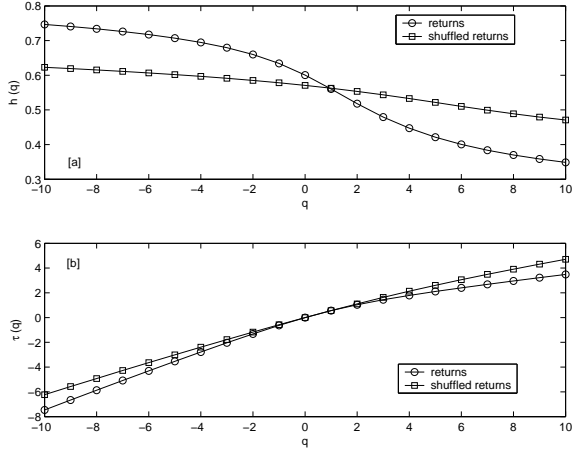}
\caption{[a](NASDAQ composite index) Scaling exponents h(q) values
for various $q$ values and [b] $\tau(q)$ representation of h(q)
values for various $q$ values, where $\tau(q) = qh(q)$.}
\end{figure}

\begin{figure}
\centering
\includegraphics[width=3in]{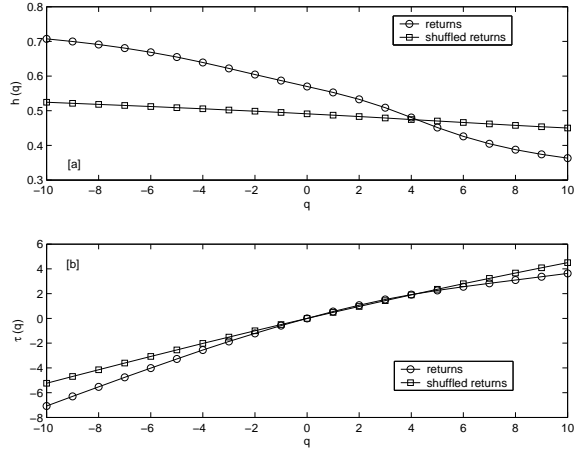}
\caption{[a] (BSE sensex) Scaling exponents $h(q)$ values for
various $q$ values and [b] $\tau(q)$ representation of h(q) values
for various $q$ values, where $\tau(q) = qh(q)$.}
\end{figure}

The scaling behaviour of the observed data sets can also be studied
by evaluating $f(\alpha)$ spectrum. $f(\alpha)$ values are obtained
from Legendre transform of $\tau(q)$: $ f(\alpha) \equiv q \alpha -
\tau(q)$, where $\alpha \equiv \frac{d\tau(q)}{dq}$. For monofractal
time series, $\alpha=const.$, whereas for multifractal time series
there occurs a distribution of $\alpha$ values. The $f(\alpha)$
spectra for the two time series are shown in Figs. 7 and 8. For the
unshuffled returns, one observes a broader spectrum, whereas for the
shuffled returns, where the correlation is lost, the same is
narrower.

\begin{figure}
\centering
\includegraphics[width=3in]{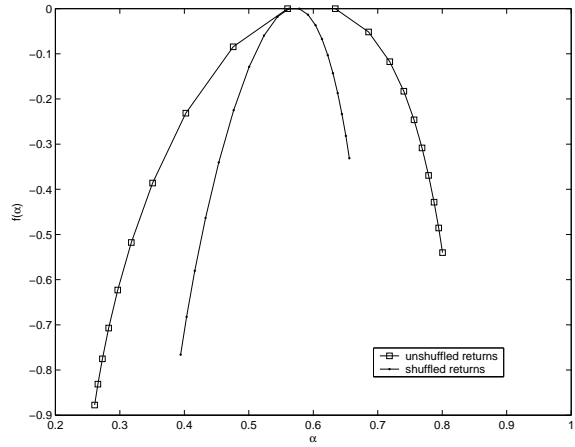}
\caption{From the integrated returns of NASDAQ composite index
values, the calculated multifractal spectrum is broader than the
spectrum of shuffled returns.}
\end{figure}

\begin{figure}
\centering
\includegraphics[width=3in]{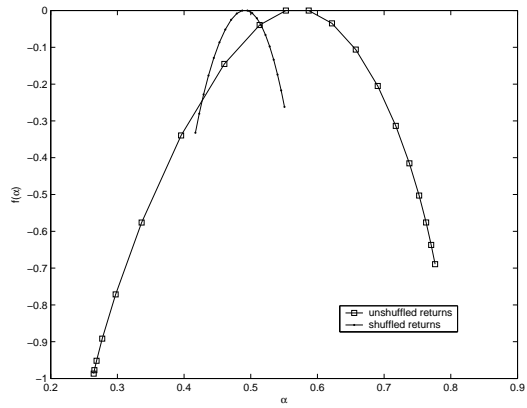}
\caption{From the integrated returns of BSE sensex values, the
calculated multifractal spectrum is much broader than the spectrum
of shuffled returns, when compared to the NASDAQ composite index
values.}
\end{figure}

The semi-log plot of distribution of logarithmic returns of NASDAQ
composite index values is shown in Fig. 9. It exhibits fat tails and
non-Gaussian features. In case of BSE sensex, the semi-log plot of
distribution of logarithmic returns shows fat tails, which are less
prominent, as shown in Fig. 10. The distribution is quite similar to
Gaussian white noise, revealing distinct differences between NASDAQ
composite index and BSE sensex values. Although correlation is
present in the two time series, they reveal distinct probability
distributions once the correlation is removed. Furthermore, we
observer that in case of BSE sensex values, the calculated
multifractal spectrum is much broader than the spectrum of shuffled
returns, when compared to the NASDAQ composite index values.

\section{Conclusion}
In conclusion, the wavelet based method presented here is found to
be quite efficient in extracting fluctuations from trend. It reveals
the distinct differences in the long-range correlation, as well as
fractal behavior of the two stock index values. Strong
non-statistical correlation is observed in BSE index, whereas the
NASDAQ index showed a multifractal behavior with long-range
statistical correlations. In case of BSE index the removal of
correlation revealed the Gaussian random noise character of the
fluctuations. It is interesting to note that the effect of country
specific parameters like corruption on economic development and
investment has been recently quantified through scaling analysis
\cite{shao}. In a similar manner, the above observed differences
between the two stock indices belonging to two different economic
environment is probably due to local dynamics.

\begin{figure}
\centering
\includegraphics[width=3in]{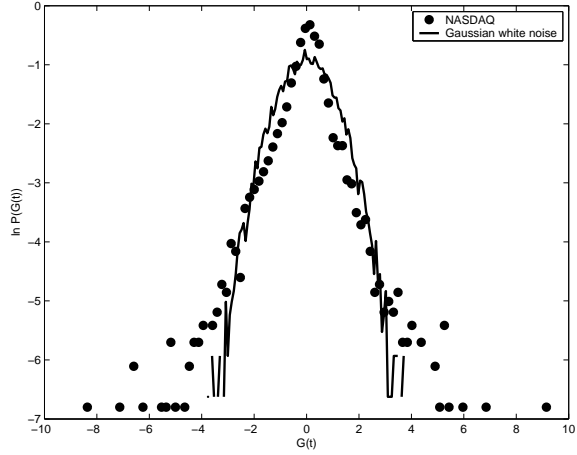}
\caption{The semi-log plot of distribution of logarithmic returns of
NASDAQ composite index values compared with Gaussian distribution.}
\end{figure}

\begin{figure}
\centering
\includegraphics[width=3in]{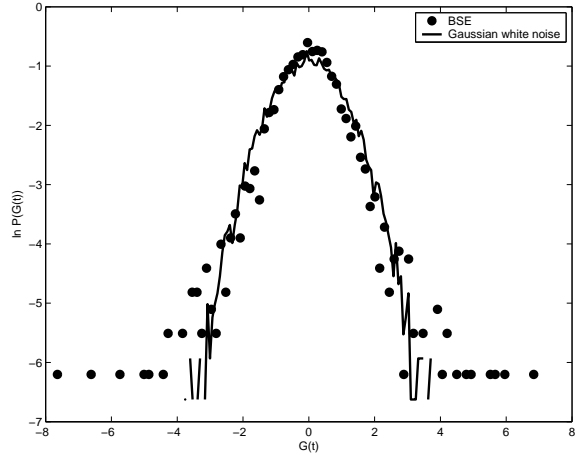}
\caption{The semi-log plot of distribution of logarithmic returns
BSE sensex values compared with Gaussian distribution.}
\end{figure}

\end{document}